\documentclass[runningheads]{llncs}

\usepackage[T1]{fontenc}
\usepackage{graphicx}
\usepackage{subcaption}
\usepackage{booktabs}
\usepackage{multirow}
\usepackage{amsmath}
\usepackage{amssymb}

\begin{document}

\title{Signed-Graph Recommendation as Structural Consistency Maximization}

\author{Zifan Wang\ \and 
 Siyu Chen\and
Wenzhuo Song\thanks{Corresponding author. Email: wzsong@nenu.edu.cn}}
% \author{Anonymous Author(s)}
% \authorrunning{Wang et al.}
\institute{}
\institute{Northeast Normal University, Changchun, Jilin, China}

\maketitle

\begin{abstract}
While signed social recommendation has shown great potential by modeling both trust and distrust relations, its effectiveness is often hindered by structural noise and data sparsity. In this work, we first identify a fundamental inconsistency across the structural, propagation, and semantic layers of existing models, which leads to biased representations learned from sparse or noisy datasets. Furthermore, we observe that most existing methods treat the observed graph as fixed, failing to bridge the gap between noisy topologies and reliable social semantics. To address these issues, we propose a unified framework named SSC-Loop that treats signed social recommendation as the maximization of structural consistency. SSC-Loop includes three dedicated modules: ESA-DA for structural consistency, a P/N/O propagation mechanism for propagation consistency, and a contrastive learning objective for semantic consistency. Experiments on Epinions demonstrate that SSC-Loop achieves strong performance on explicit signed social rating prediction, while auxiliary results on Slashdot under a derived link-existence setting further suggest its ability to exploit signed social structures. Source code is available at 
\url{https://github.com/Refrainwww/SSC-Loop}.

\end{abstract}

\section{Introduction}

Social recommendation leverages user--item interactions and social relations to improve personalized recommendation
~\cite{tang2013social,ma2011recommender}. 
Most existing methods assume that social relations are homogeneous and positive, such as friendship or trust~\cite{fan2019graph}. However, real-world social networks often contain both trust and distrust relations~\cite{leskovec2010predicting}. Signed social recommendation is therefore important, as it can incorporate both trust and distrust relations into user preference modeling~\cite{wang2017signed,huang2019signed}.

Despite this potential, signed social recommendation remains challenging because observed signed graphs are often sparse, noisy, and structurally imbalanced. Recent works commonly adopt Graph Neural Networks (GNNs), which model social relationships via message passing over observed links~\cite{fan2019graph}. However, unreliable topology can distort information propagation and induce biased representations. In benchmark signed networks such as Epinions and Slashdot, social links are highly sparse~\cite{massa2007trust}, and many signed triangles may violate structural balance theory~\cite{heider1946attitudes}. These issues reveal a key limitation of existing methods, i.e., graph structure, propagation dynamics, and learned representations may be mutually inconsistent.

In this work, we characterize this problem as a mismatch among three coupled layers: observed signed topology, polarity-aware propagation dynamics, and learned representation geometry. Structural inconsistency indicates that the observed topology may not reflect reliable social semantics. Propagation inconsistency indicates that positive and negative signals may be weakened, mixed, or distorted during multi-hop message passing. Semantic inconsistency indicates that the learned embedding space may fail to place trusted users close to each other and distrusted users far apart.

Based on this perspective, we formulate signed social recommendation as a structural consistency maximization problem. Rather than treating the observed graph as a fixed input, we argue that graph structure and user representations should co-evolve within a closed loop. To this end, we propose the Signed Structural Consistency Loop (SSC-Loop), which integrates a module named ESA-DA for adaptive signed graph refinement, a polarity-aware P/N/O aggregation scheme for polarity-preserving propagation, and contrastive relation alignment for sign-aware semantic learning. In SSC-Loop, representations improved in each iteration enable more reliable graph refinement, while the refined graph further improves subsequent propagation and representation learning.

Our contributions are summarized as follows:
\begin{enumerate} 
\item We identify a three-level inconsistency problem in signed social recommendation, including structural, propagation, and semantic inconsistency. 
\item We formulate signed social recommendation as a structural consistency maximization problem and propose a co-evolutionary framework for joint graph and representation refinement. 
\item We design ESA-DA, P/N/O propagation, and contrastive relation alignment to improve structural reliability, polarity-preserving propagation, and signed semantic alignment, respectively. 
\item Experiments on Epinions and auxiliary results on Slashdot under a link-existence setting demonstrate the effectiveness and robustness of SSC-Loop. 
\end{enumerate}

\section{Related Work}

Social recommendation extends traditional collaborative filtering by exploiting social relations as auxiliary information. Early methods such as SocialMF~\cite{jamali2010matrix} and TrustMF~\cite{yang2016trustmf} regularize user representations based on the assumption that socially connected users tend to have similar preferences. However, this assumption is limited in signed networks, where distrust relations may also exist. Signed social recommendation methods such as TDRec~\cite{bai2015tdrec} and RecSSN~\cite{tang2016recommendations} explicitly model trust and distrust by encouraging trusted users to have similar representations while maximizing the difference between distrusted users. Signed network representation learning has also been studied from node/edge embedding and non-Euclidean geometry perspectives~\cite{song2018nodeedge,song2021hyperbolic}.

Graph-based recommendation models further improve high-order information propagation. LightGCN~\cite{he2020lightgcn} simplifies graph convolution for collaborative filtering, GraphRec~\cite{fan2019graph} incorporates social neighborhoods into the recommendation process, and SIGformer~\cite{chen2024sigformer} introduces a sign-aware Transformer-based propagation mechanism. Beyond recommendation-specific models, signed network structure has also been analyzed through block modeling and semi-supervised stochastic block models, which aim to reveal latent community patterns and structural regularities in networks with positive and negative links~\cite{zhao2018block,song2020ssbm,liu2025sbmsurvey}. 

However, most existing recommendation methods treat the observed social graph as fixed and still decouple the three levels of consistency: the graph structure is rarely optimized to support reliable signed information propagation; the propagation process lacks explicit constraints to preserve polarity-consistent semantics; and the learned representations cannot be jointly aligned with structural patterns and propagation characteristics. In sparse, imbalanced, and noisy signed social networks, such cross-level inconsistencies may accumulate and substantially degrade recommendation performance~\cite{wang2026affectagent,wang2026emotiontree,zhao2026phase,zhao2026affectverse}.

\section{The Proposed Method}
\label{sec:method}

We consider a signed social recommendation setting with a user set $\mathcal{U}$ and an item set $\mathcal{I}$, where $N_u=|\mathcal{U}|$ and $N_i=|\mathcal{I}|$ denote the numbers of users and items, respectively. The signed social network is denoted by $\mathcal{G}_S=(\mathcal{U},\mathcal{E}_S)$, where $\mathcal{U}$ is the node set and $\mathcal{E}_S\subseteq \mathcal{U}\times\mathcal{U}\times\{+1,-1\}$ is the set of signed user--user relations. Each element $(u,v,s_{uv})\in\mathcal{E}_S$ indicates that user $u$ has a signed social relation to user $v$, where $s_{uv}=+1$ denotes a positive relation, such as trust, and $s_{uv}=-1$ denotes a negative relation, such as distrust.

Let $\mathbf{R}\in\mathbb{R}^{N_u\times N_i}$ denote the user--item interaction matrix, where $R_{ui}$ is user $u$'s observed preference for item $i$ when available. Signed social recommendation aims to predict user $u$'s unknown preference $\hat{R}_{ui}$ for item $i$ by jointly exploiting user--item interactions and signed user--user relations.

Reliable signed recommendation requires graph structure, message propagation, and representation semantics to be jointly consistent. Instead of treating the observed signed graph as fixed, we formulate signed social recommendation as consistency maximization over structural, propagation, and semantic levels, and instantiate it through the proposed SSC-Loop framework.

\subsection{Consistency Formulation}

\paragraph{Structural Consistency.}
Structural consistency measures the extent to which the signed graph provides a reliable basis for recommendation under signed social priors. A structurally consistent graph should satisfy structural balance while preserving reasonable connectivity, avoiding pathological patterns such as hub explosion and isolated nodes. Improving structural consistency thus requires removing unreliable links and recovering reliable missing relations under explicit structural constraints.

\paragraph{Propagation Consistency.}
Propagation consistency measures how well polarity-aware information persists during multi-hop message passing. In signed recommendation, positive, negative, and neutral signals should not mix indiscriminately, as such mixing may cause semantic dilution, cancellation, or unintended sign inversion. A propagation-consistent model should preserve polarity across hops while capturing higher-order relational dependencies.

\paragraph{Semantic Consistency.}
Semantic consistency measures how well the geometry of the learned embedding space matches signed link polarity. Users connected by positive edges should lie closer in representation space than users connected by negative edges, so that trust corresponds to proximity and distrust corresponds to separation. Enforcing semantic consistency helps prevent ambiguity in the learned representations for signed relations.

\begin{figure*}[ht]
    \centering
    \includegraphics[width=\textwidth]{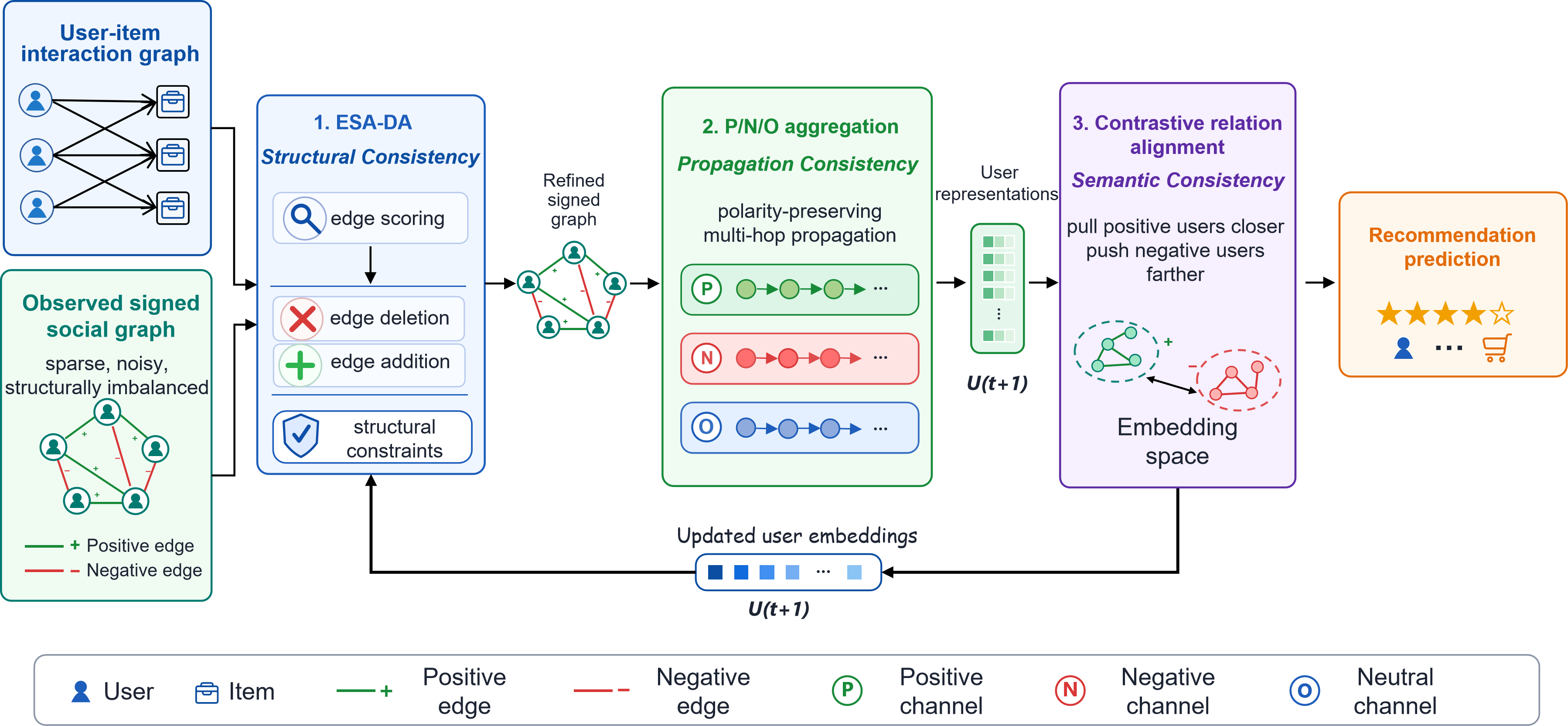}
    \caption{The overall architecture of SSC-Loop.}
    \label{fig:ssc-loop-framework}
\end{figure*}

As illustrated in Fig.~\ref{fig:ssc-loop-framework}, SSC-Loop forms a closed-loop framework in which graph topology and learned representations are iteratively refined for higher consistency. ESA-DA improves structural consistency by estimating edge confidence from the current model state and editing the signed graph under degree, connectivity, and balance constraints. Using the refined graph, the P/N/O aggregation module improves propagation consistency by separating positive, negative, and neutral signals into distinct channels, preserving polarity-aware information during multi-hop propagation. Then, contrastive relation alignment improves semantic consistency by explicitly shaping the embedding geometry to pull positively linked users closer and push negatively linked users farther apart. The updated representations are fed back to the structural refinement module, forming a consistency feedback loop that jointly aligns graph structure, message propagation, and representation semantics.

\subsection{Structural Consistency: ESA-DA}

The goal of ESA-DA is to refine the graph by removing unreliable edges and adding reliable ones via a unified polarity-aware scoring and selection framework.

\paragraph{Confidence Scoring with Item-Conditioned Context.}

We first define an edge compatibility score $r_{uv}$ that measures the signed affinity between two users under the current model state.

Instead of relying solely on direct embedding similarity, we incorporate item-conditioned contextual information. Let $e_u$, $e_v$, and $z_i$ denote the embeddings of user $u$, user $v$, and item $i$, respectively. Let $\mathcal{I}(u)$ be the set of items interacted with by user $u$. We compute:
\begin{equation}
\alpha_{ui}^{(v)}
=
\frac{
\exp(\mathbf{z}_i^\top \mathbf{e}_v)
}{
\sum_{j \in \mathcal{I}(u)} \exp(\mathbf{z}_j^\top \mathbf{e}_v)
},
\quad i \in \mathcal{I}(u),
\end{equation}
\begin{equation}
\mathbf{c}_{v|u}
=
\sum_{i \in \mathcal{I}(u)}
\alpha_{ui}^{(v)} \mathbf{z}_i .
\end{equation}
If $\mathcal{I}(u)$ is empty, we set $\mathbf{c}_{v|u}=\mathbf{e}_v$. The edge compatibility score is then:
\begin{equation}
{r}_{uv}
=
\cos(\mathbf{e}_u, \mathbf{c}_{v|u}).
\end{equation}
We map this score into polarity-aware scores via a temperature-scaled sigmoid:
\begin{equation}
p_{uv}^{+}
=
\sigma\left(\frac{r_{uv}}{\tau}\right),
\quad
p_{uv}^{-}
=
\sigma\left(-\frac{r_{uv}}{\tau}\right),
\end{equation}
where $\tau$ denotes the temperature parameter. A larger $p_{uv}^{+}$ indicates stronger evidence for a positive relation, while a larger $p_{uv}^{-}$ indicates stronger evidence for a negative relation.
These values are not interpreted as calibrated probabilities, but as scores used for ranking and selection.

\paragraph{Candidate Selection for Edge Editing.}

Using the unified scoring function, ESA-DA performs both edge deletion and edge addition.

For observed edges, ESA-DA first converts the signed graph into a unique edge set to avoid duplicated edge evaluation. Positive observed edges are ranked in ascending order of $p_{uv}^{+}$, and negative observed edges are ranked in ascending order of $p_{uv}^{-}$. Edges with the lowest confidence are treated as deletion candidates, since they are least consistent with the current model state.

For unobserved relations, ESA-DA avoids enumerating all possible non-edges, which is computationally expensive. Therefore, we restrict candidate generation to each user's $k$-nearest-neighbor set in the normalized embedding space. Let $\bar{\mathbf{e}}_u = \mathbf{e}_u / \|\mathbf{e}_u\|_2$ be the normalized user embedding. For an unobserved pair $(u,v)$ with $v \in \mathrm{kNN}(u)$ and $(u,v)$ absent from the graph, ESA-DA computes
\begin{equation}
\tilde{r}_{uv}
=
\bar{\mathbf{e}}_u^\top \bar{\mathbf{e}}_v.
\end{equation}

We then reuse the same polarity mapping:
\begin{equation}
q_{uv}^{+}
=
\sigma\left(\frac{\tilde{r}_{uv}}{\tau}\right),
\quad
q_{uv}^{-}
=
\sigma\left(-\frac{\tilde{r}_{uv}}{\tau}\right).
\end{equation}
The scores $q_{uv}^{+}$ and $q_{uv}^{-}$ are used to maintain two candidate pools for positive and negative edge additions, respectively. Only the top-$K_c$ candidates for each sign are retained, which makes graph expansion efficient while focusing on the most confident missing relations.

\paragraph{Structure Guards.}
High-confidence candidates are not always safe, since unconstrained graph editing can create hubs, isolate nodes, or worsen signed imbalance. We therefore impose three safeguards. Degree caps $d_{\text{max}}^{+}$ and $d_{\text{max}}^{-}$ bound the maximum positive and negative degrees to prevent hub explosion. A connectivity guard ensures that edge deletion does not isolate nodes, i.e., each endpoint must keep degree at least $\delta_{\text{min}}$ after deletion. A balance guard accepts a candidate addition $(u,v)$ with sign $s \in \{+1,-1\}$ only when it does not increase structural imbalance.

Let $\mathbf{A}_{\mathrm{abs}}$ and $\mathbf{A}_{\mathrm{sgn}}$ denote the unsigned and signed adjacency matrices, respectively. The two-hop unsigned and signed co-occurrence scores between $u$ and $v$ are computed as:
\begin{equation}
M_{\mathrm{abs}}^{(2)}[u,v]
=
\mathbf{A}_{\mathrm{abs}}^2[u,v],
\quad
M_{\mathrm{sgn}}^{(2)}[u,v]
=
\mathbf{A}_{\mathrm{sgn}}^2[u,v].
\end{equation}
ESA-DA then evaluates
\begin{equation}
\Delta_{\mathrm{neg}}
=
M_{\mathrm{abs}}^{(2)}[u,v]
-
s \cdot M_{\mathrm{sgn}}^{(2)}[u,v],
\end{equation}
and accepts the candidate only when $\Delta_{\mathrm{neg}}$ is non-positive.
This criterion favors edits that are compatible with signed structural balance, such as completing balanced motifs, and suppresses additions that intensify inconsistent two-hop signed patterns.

After edge deletion and addition, the edited edges form a symmetric refined signed adjacency matrix, which replaces the original social graph. ESA-DA does not directly modify user embeddings; instead, the refined graph updates model parameters and user representations in the next training loop.

\subsection{Propagation Consistency: Signed High-order Triplet Aggregation (P/N/O)}

Propagation consistency aims to preserve polarity-aware information during message passing. We propose a P/N/O aggregation mechanism that separates positive, negative, and neutral signals to avoid their uncontrolled mixing.
\paragraph{User--Item Convolution Backbone.}
Before propagating signed social signals, the model derives interaction-aware seed representations from the user--item graph. SSC-Loop applies mean aggregation on the interaction graph to obtain the initial fused embeddings:
\begin{equation}
\mathbf{E} = \left[ \mathbf{h}^{(0)}_u \,\|\, \tilde{\mathbf{A}}_{ui}\,\mathbf{h}^{(0)}_i \right].
\end{equation}
This step yields interaction-structure-aware embeddings that do not yet encode polarity, providing the basis for subsequent signed propagation.

\paragraph{Sign and Similarity-Aware Edge Splitting.}
To avoid mixing heterogeneous signals in a single stream, we divide signed relations into Positive (P), Negative (N), and Neutral (O) channels based jointly on link polarity and embedding similarity. Edges whose sign agrees with strong embedding similarity are routed to the P or N channel, whereas weakly consistent or ambiguous relations are assigned to the neutral channel. Specifically, positively signed and semantically similar pairs go to P, negatively signed and semantically dissimilar pairs to N, and edges with weak similarity evidence to O.

\paragraph{High-Order P/N/O Propagation.}
The key challenge of signed propagation is to preserve polarity semantics while capturing higher-order signed dependencies. We maintain three channel-specific user representation matrices at layer $l$, denoted by $\mathbf{P}^{(l)}, \mathbf{N}^{(l)}, \mathbf{O}^{(l)} \in \mathbb{R}^{N_u \times d}$, corresponding to the positive, negative, and neutral channels, respectively. Moreover, we denote by $\mathrm{PCM}$, $\mathrm{NCM}$, and $\mathrm{OCM}$ the positive-channel, negative-channel, and neutral-channel normalized propagation matrices, respectively. Here, each propagation matrix is constructed from the adjacency structure of its corresponding channel and normalized so that message aggregation remains scale-consistent across users with different degrees.

Based on these channel-specific propagation operators, the positive and negative channels are updated as
\begin{equation}
\begin{aligned}
\mathbf{P}^{(l)} &= \tfrac{1}{2}\left(\mathrm{PCM}\,\mathbf{P}^{(l-1)} + \mathrm{NCM}\,\mathbf{N}^{(l-1)}\right),\\
\mathbf{N}^{(l)} &= \tfrac{1}{2}\left(\mathrm{NCM}\,\mathbf{P}^{(l-1)} + \mathrm{PCM}\,\mathbf{N}^{(l-1)}\right),\\
\mathbf{O}^{(l)} &= \text{Aggregated from P, N, O interactions}.
\end{aligned}
\end{equation}
This mechanism explicitly preserves channel-specific semantics and models signed multi-hop logic through controlled interactions among the P, N, and O channels.

After propagation, the final user representation is denoted by $\mathbf{e}_u$, which is used for both recommendation and semantic alignment.

\subsection{Semantic Consistency: Contrastive Relation Alignment}

% \paragraph{Objective.}
To enforce semantic consistency, we introduce an intra-view signed contrastive loss that pulls positive pairs closer and pushes negative pairs farther apart:
\begin{equation}
\mathcal{L}_{\text{CL}}=
\begin{cases}
1-c, & s=+1 \quad (\text{pull}),\\[2pt]
\max(0,\;c+m), & s=-1 \quad (\text{push}),
\end{cases}
\end{equation}
where $s$ is the sign of the user--user relation, $c=\cos(\mathbf{e}_u,\mathbf{e}_v)$, and $m>0$ is a margin hyperparameter. The overall objective is
\begin{equation}
\mathcal{L}=(1-\alpha)\,\mathcal{L}_{rec}+\alpha\,\mathcal{L}_{\text{CL}},
\end{equation}
where $\mathcal{L}_{rec}$ denotes the recommendation loss and $\alpha$ controls the trade-off between recommendation accuracy and semantic alignment. This objective aligns representation geometry with link polarity and makes the learned embeddings more discriminative for recommendation.

\subsection{Consistency Feedback Loop}

SSC-Loop lets signed graph structure and user representations co-evolve. Current representations refine the graph by removing unreliable or structurally implausible links and recovering reliable missing relations satisfying signed balance and connectivity constraints. Propagation on this refined graph better preserves polarity information and yields higher-quality representations.

To realize this bidirectional coupling, we adopt an EM-style alternating refinement strategy. With the model parameters fixed, current user embeddings score signed edges, generate candidates for edge addition and deletion, and refine the graph through ESA-DA under degree, connectivity, and balance constraints. With the refined graph fixed, polarity-preserving propagation and semantic alignment are then performed on the updated graph, and the model parameters are optimized by minimizing the objective function $L$. This yields more informative and discriminative user embeddings, which are further used for graph refinement in the next outer-loop iteration.

The procedure is repeated for up to $K$ outer-loop iterations, allowing the topology and the representations to progressively improve their structural, propagation, and semantic consistency. Because graph refinement involves discrete edge editing under multiple structural constraints, SSC-Loop does not rely on a strict theoretical guarantee of monotonic convergence over a fixed continuous objective. Instead, we use a validation-based stopping strategy in practice. Specifically, after each outer-loop iteration, we evaluate the model on the validation set and select the checkpoint with the best validation performance. Iteration stops when the validation metric no longer improves for a predefined patience window or when the maximum number of outer-loop iterations $K$ is reached.

\section{Experiments}

\subsection{Datasets}
\label{sec:dataset}

We use Epinions as the main benchmark for evaluating rating prediction in signed social recommendation, because it provides both explicit trust/distrust relations and user--item ratings. To the best of our knowledge, it is the only publicly available dataset that supports this setting. To further examine whether SSC-Loop can exploit signed social structures beyond explicit rating prediction, we also report auxiliary results on Slashdot. Since Slashdot contains friend/foe relations but no real user--item ratings, we construct a pseudo-recommendation task from its signed user--user network. Specifically, the same set of Slashdot users is used on both sides of the derived interaction matrix: a source-side node is treated as a user, and a target-side node is treated as a pseudo-item. A directed social link from user $u_i$ to user $u_j$ is converted into a binary interaction label $R_{ij}=1$, while sampled unlinked user pairs are used as zero-valued entries. The friend/foe polarity is not used as the prediction target in $\mathbf{R}$; instead, it is retained in the signed social graph $\mathcal{G}_S$ for signed propagation and structural modeling.

The statistics reported in Table~\ref{tab:dataset-stats} are obtained after preprocessing.

\begin{table*}[t]
\centering
\caption{Statistics of the datasets used in our experiments.}
\label{tab:dataset-stats}
\begin{tabular}{lcc}
\toprule
\textbf{Statistic} & \textbf{Epinions} & \textbf{Slashdot} \\
\midrule
Dataset type & Signed social recommendation & Signed social network \\
Users / Nodes & 180{,}202 & 82{,}168 \\
Items & 755{,}761 & 82{,}168 \\
Ratings / Interactions & 13{,}668{,}320 & 549{,}202 \\
Social links / Edges & 841{,}372 & 549{,}202 \\
Prediction target & Explicit rating (1--5) & Derived binary link label (0/1) \\

Rating sparsity & 99.98\% & 99.99\% \\
\bottomrule
\end{tabular}
\end{table*}

\subsection{Experimental Setting}

We evaluate SSC-Loop on Epinions as the primary benchmark and report auxiliary results on Slashdot for signed-structure validation. For Epinions, the observed user--item rating entries are randomly divided into training, validation, and test sets. Specifically, we use 70\% of the observed entries for training, 20\% for testing, and 10\% samples as the validation set for early stopping, checkpoint selection, and hyperparameter tuning. To reduce the influence of randomness, each experiment is repeated five times with different random seeds, and the average result is reported. All methods are evaluated under the same preprocessing pipeline, data split protocol, and metric setting. 

For Slashdot, the evaluation is conducted on a binary link-existence prediction task. We first split the observed directed friend/foe links into training, validation, and test links according to the same split ratio. Only the training links are used to construct the signed social graph $\mathcal{G}_S^{train}$ and the positive entries in the training interaction matrix $\mathbf{R}^{train}$. Validation and test links are removed from $\mathcal{G}_S^{train}$ before model training and are used only as held-out positive labels for model selection and final evaluation, respectively. Therefore, the model input does not contain the existence or polarity information of validation or test links.

Zero-valued entries in Slashdot are constructed by sampling unlinked user pairs that do not correspond to any observed friend/foe link in the original Slashdot network. Negative samples are generated separately for training, validation, and testing, and the validation/test negative samples are not used during model training. The polarity of friend and foe relations is used only in the signed social graph for propagation and structural modeling, whereas the prediction target in the derived matrix is binary: linked pairs are labeled as $1$, and sampled unlinked pairs are labeled as $0$. Since the evaluated models output real-valued prediction scores and no clipping operation is applied to constrain the predictions to $[0,1]$, RMSE and MAE on Slashdot may exceed 1 even though the ground-truth labels are binary. Therefore, RMSE and MAE on Epinions measure explicit rating prediction error, whereas RMSE and MAE on Slashdot measure real-valued reconstruction error on the derived link-existence prediction task.

We compare SSC-Loop with representative baselines from three categories. Specifically, SocialMF~\cite{jamali2010matrix} and TrustMF~\cite{yang2016trustmf} are used as social recommendation baselines; TDRec~\cite{bai2015tdrec} and RecSSN~\cite{tang2016recommendations} are used as signed social recommendation baselines; and LightGCN~\cite{he2020lightgcn}, GraphRec~\cite{fan2019graph}, and SIGformer~\cite{chen2024sigformer} are used as graph-based recommendation baselines.

\begin{table*}[t]
\centering
\caption{Performance comparison on the Epinions and Slashdot datasets. The best results are shown in \textbf{bold} and the second-best are \underline{underlined}.}
\label{tab:comparison}
\begin{tabular}{p{7cm}cccc}
\toprule
\multirow{2}{*}{\textbf{Model}} & \multicolumn{2}{c}{\textbf{Epinions}} & \multicolumn{2}{c}{\textbf{Slashdot}} \\
\cmidrule(lr){2-3} \cmidrule(lr){4-5}
& \textbf{RMSE} & \textbf{MAE} & \textbf{RMSE} & \textbf{MAE} \\
\midrule
SocialMF & 0.5891 & 0.3311 & 1.7546 & 1.5701 \\
TrustMF & 0.5563 & 0.3250 & 1.6226 & 1.3201 \\
TDRec & 0.5263 & 0.3375 & 1.9941 & 1.6327 \\
RecSSN & 0.5667 & 0.3198 & 1.8805 & 1.4317 \\
LightGCN & 0.5698 & 0.3470 & 1.3921 & 0.9431 \\
GraphRec & 0.5153 & \underline{0.3036} & 1.2564 & \textbf{0.7318} \\
SIGformer & \underline{0.4658} & 0.3090 & \underline{1.0929} & 0.7737 \\
\textbf{SSC-Loop (Ours)} & \textbf{0.4398} & \textbf{0.2489} & \textbf{1.0650} & \underline{0.7694} \\
\bottomrule
\end{tabular}
\end{table*}

\begin{table}[t]
\centering
\caption{Ablation study on Epinions.}
\label{tab:ablation}
\begin{tabular}{p{9cm}cc}
\toprule
\textbf{Variant} & \textbf{RMSE} & \textbf{MAE} \\
\midrule
w/o ESA-DA & 0.4838 & 0.2984 \\
w/o P/N/O & 0.4761 & 0.2981 \\
w/o Contrastive & 0.4896 & 0.2790 \\
\textbf{Full SSC-Loop} & \textbf{0.4398} & \textbf{0.2489} \\
\bottomrule
\end{tabular}
\end{table}

\subsection{Performance Analysis}

Table~\ref{tab:comparison} reports the overall comparison on Epinions and Slashdot. On Epinions, SSC-Loop achieves the best RMSE and MAE among all compared methods, reducing RMSE from 0.4658 to 0.4398 compared with the strongest baseline SIGformer. This result shows the benefit of jointly refining signed topology, preserving polarity-aware propagation, and enforcing signed semantic alignment.

On Slashdot, SSC-Loop obtains the best RMSE and the second-best MAE under the derived binary interaction setting. Since Slashdot does not contain real item ratings, these results are used only as auxiliary evidence that SSC-Loop can exploit signed social structures beyond explicit rating prediction.

\subsection{Ablation Study and Further Analysis of SSC-Loop}
Table~\ref{tab:ablation} shows that any component removal degrades performance on Epinions. Removing ESA-DA causes the largest drop, highlighting the importance of adaptive topology refinement. Removing P/N/O weakens polarity-preserving propagation, while removing the contrastive objective reduces signed semantic alignment. Together, these modules provide complementary benefits.

\begin{figure*}[t]
    \centering
    \begin{subfigure}[t]{0.32\textwidth}
        \vspace{0pt}
        \centering
        \includegraphics[width=\textwidth]{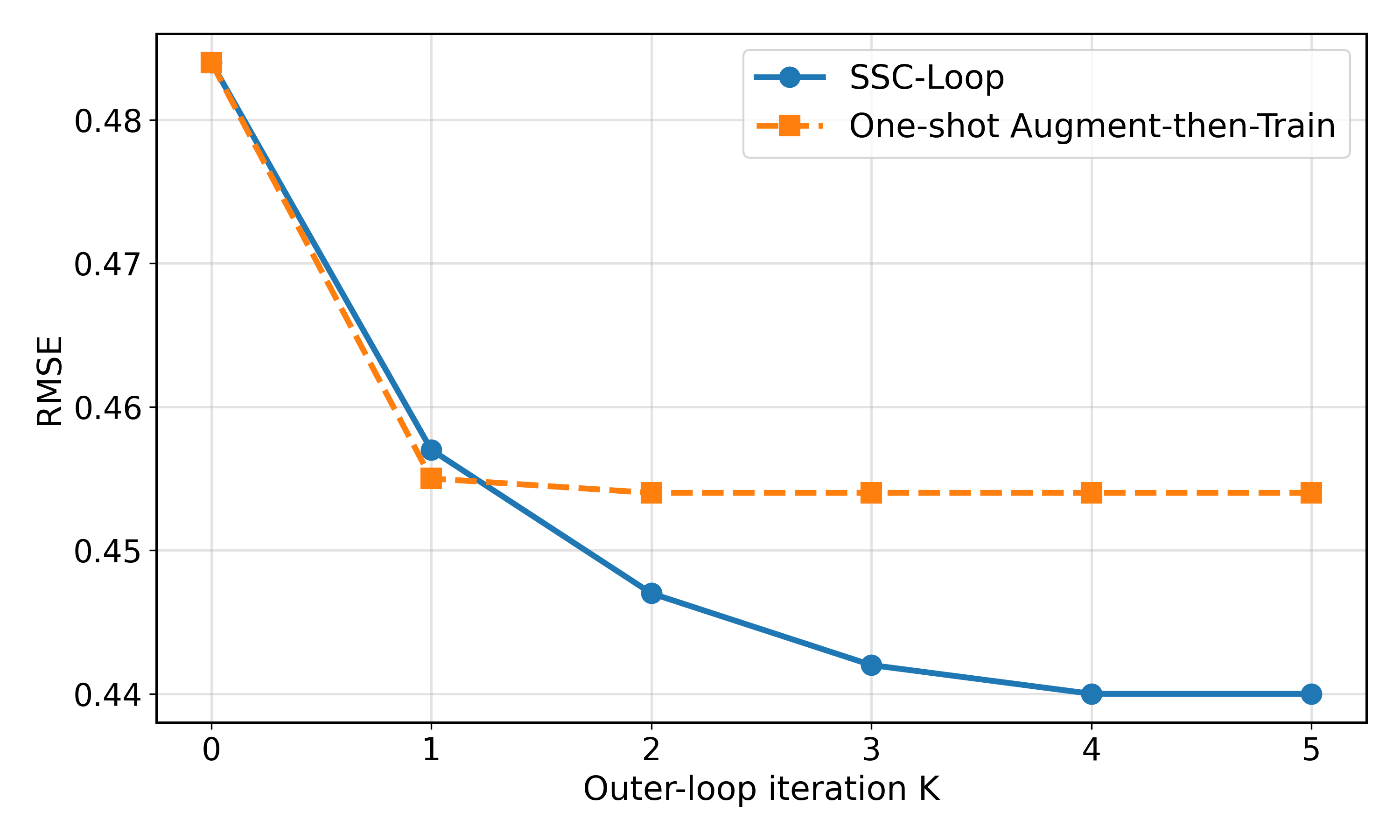}
        \caption{Closed-loop iteration}
        \label{fig:loop-iteration}
    \end{subfigure}
    \hfill
    \begin{subfigure}[t]{0.32\textwidth}
        \vspace{0pt}
        \centering
        \includegraphics[width=\textwidth]{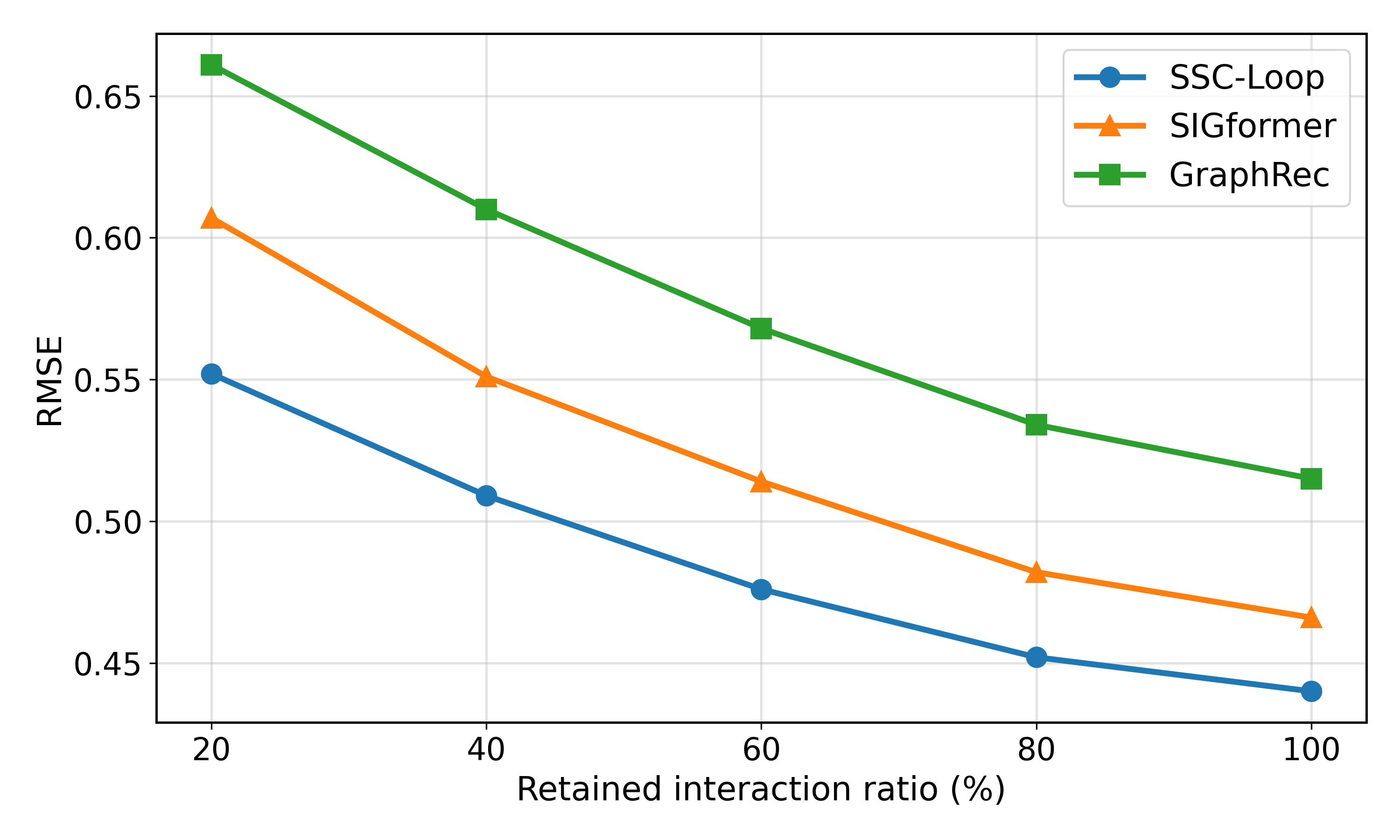}
        \caption{Sparsity robustness}
        \label{fig:sparsity-robustness}
    \end{subfigure}
    \hfill
    \begin{subfigure}[t]{0.32\textwidth}
        \vspace{0pt}
        \centering
        \includegraphics[width=\textwidth]{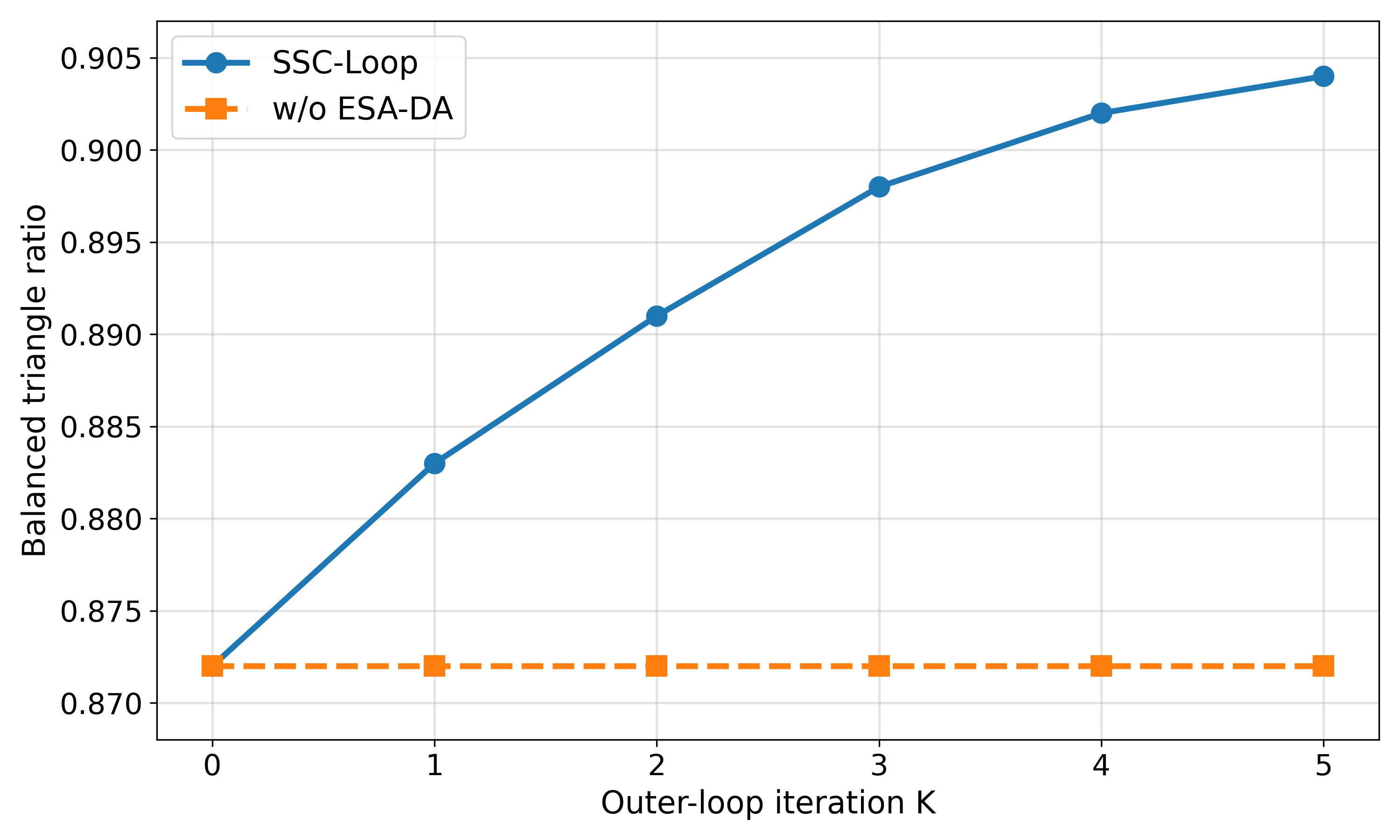}
        \caption{Structural balance trend}
        \label{fig:balance-ratio}
    \end{subfigure}

    \caption{Detailed analysis of SSC-Loop.}

    \label{fig:combined_analysis}
\end{figure*}

We further analyze SSC-Loop from three perspectives: iterative feedback, sparsity robustness, and structural balance. First, Fig.~\ref{fig:loop-iteration} shows that closed-loop feedback reduces RMSE from 0.4842 at $K=0$ to 0.4398 at $K=4$, while the one-shot augment-then-train variant saturates around 0.4540, indicating that iterative co-evolution between topology refinement and representation learning is more effective than static augmentation. Second, under sparse rating supervision, we retain 20\%, 40\%, 60\%, 80\%, and 100\% of training interactions while keeping the test set fixed. As shown in Fig.~\ref{fig:sparsity-robustness}, SSC-Loop consistently outperforms GraphRec and SIGformer in RMSE, showing stronger ability to exploit signed social information with limited interactions. Finally, based on structural balance theory, Fig.~\ref{fig:balance-ratio} shows that SSC-Loop increases the balanced triangle ratio from 0.8721 to 0.9040, whereas the variant without ESA-DA remains nearly unchanged. This suggests that ESA-DA refines the signed graph in a structurally meaningful way and better supports propagation.

\section{Conclusion}

In this paper, we proposed SSC-Loop, a closed-loop framework for signed social recommendation. It combines ESA-DA for adaptive signed graph refinement, P/N/O propagation for polarity-aware high-order message passing, and contrastive relation alignment for sign-aware semantic learning. Experiments on Epinions show improved signed rating prediction, while results on Slashdot demonstrate its effectiveness in exploiting signed structures for derived link-existence prediction. Further analyses confirm the contribution of each component, the value of iterative feedback, robustness under sparse supervision, and improved structural balance. Future work will extend SSC-Loop to temporal, multi-relational, and multimodal signed recommendation scenarios.

\section*{Acknowledgments}

This research was funded by the National Natural Science Foundation of China under Grant 62307006, and the Fundamental Research Funds for the Central Universities.


\begin{thebibliography}{99}

\bibitem{tang2013social}
Tang, J., Hu, X., Liu, H.: Social recommendation: a review. Social Network Analysis and Mining \textbf{3}(4), 1113--1133 (2013)

\bibitem{ma2011recommender}
Ma, H., Zhou, D., Liu, C., Lyu, M.R., King, I.: Recommender systems with social regularization. In: Proceedings of the Fourth ACM International Conference on Web Search and Data Mining, pp. 287--296. ACM, New York (2011)

\bibitem{fan2019graph}
Fan, W., Ma, Y., Li, Q., He, Y., Zhao, E., Tang, J., Yin, D.: Graph neural networks for social recommendation. In: The World Wide Web Conference, pp. 417--426. ACM, New York (2019)

\bibitem{leskovec2010predicting}
Leskovec, J., Huttenlocher, D., Kleinberg, J.: Predicting positive and negative links in online social networks. In: Proceedings of the 19th International Conference on World Wide Web, pp. 641--650. ACM, New York (2010)

\bibitem{wang2017signed}
Wang, S., Tang, J., Aggarwal, C., Chang, Y., Liu, H.: Signed network embedding in social media. In: Proceedings of the 2017 SIAM International Conference on Data Mining, pp. 327--335. SIAM (2017)

\bibitem{huang2019signed}
Huang, J., Shen, H., Hou, L., Cheng, X.: Signed graph attention networks. In: International Conference on Artificial Neural Networks, pp. 566--577. Springer, Cham (2019)

\bibitem{massa2007trust}
Massa, P., Avesani, P.: Trust-aware recommender systems. In: Proceedings of the 2007 ACM Conference on Recommender Systems, pp. 17--24. ACM, New York (2007)

\bibitem{heider1946attitudes}
Heider, F.: Attitudes and cognitive organization. The Journal of Psychology \textbf{21}(1), 107--112 (1946)

\bibitem{jamali2010matrix}
Jamali, M., Ester, M.: A matrix factorization technique with trust propagation for recommendation in social networks. In: Proceedings of the Fourth ACM Conference on Recommender Systems, pp. 135--142. ACM, New York (2010)

\bibitem{wang2026affectagent}
Wang, Z., Yu, Z., Zhu, Y., Zhao, B., Liang, H., Wang, T., Xia, W.,
Zhang, J., Liu, Z., Ma, H., Ma, F., Tian, Q.:
AffectAgent: Collaborative multi-agent reasoning for retrieval-augmented multimodal emotion recognition.
arXiv preprint arXiv:2604.12735 (2026)

\bibitem{wang2026emotiontree}
Wang, Z., Zhao, B., Zhu, Y., Liu, Z., Ma, H., Zhang, R., Ding, S.,
Xie, Q., Yu, Z.:
Navigating the emotion tree: Hierarchical hyperbolic RAG for multimodal emotion recognition.
arXiv preprint arXiv:2605.18884 (2026)

\bibitem{zhao2026phase}
Zhao, B., Guo, D., Cao, J., Xu, Y., Zou, B., Tan, T., Sun, Y., Yu, Z.:
Phase-net: Physics-grounded harmonic attention system for efficient remote photoplethysmography measurement.
In: Proceedings of the IEEE/CVF Conference on Computer Vision and Pattern Recognition,
pp. 21198--21207 (2026)

\bibitem{zhao2026affectverse}
Zhao, B., Ye, F., Ji, Y., Zhao, S., Peng, X., Yu, Z.:
AffectVerse: Emotional world models for multimodal affective computing.
arXiv preprint arXiv:2605.19950 (2026)


\bibitem{yang2016trustmf}
Yang, B., Lei, Y., Liu, J., Li, W.: Social collaborative filtering by trust. IEEE Transactions on Pattern Analysis and Machine Intelligence \textbf{39}(8), 1633--1647 (2017)

\bibitem{bai2015tdrec}
Bai, T., Yang, B., Li, F.: TDRec: Enhancing social recommendation using both trust and distrust information. In: Second European Network Intelligence Conference, pp. 60--66. IEEE (2015)


\bibitem{tang2016recommendations}
Tang, J., Aggarwal, C., Liu, H.: Recommendations in signed social networks. In: Proceedings of the 25th International Conference on World Wide Web, pp. 31--40. International World Wide Web Conferences Steering Committee (2016)

\bibitem{he2020lightgcn}
He, X., Deng, K., Wang, X., Li, Y., Zhang, Y., Wang, M.: LightGCN: Simplifying and powering graph convolution network for recommendation. In: Proceedings of the 43rd International ACM SIGIR Conference on Research and Development in Information Retrieval, pp. 639--648. ACM, New York (2020)

\bibitem{chen2024sigformer}
Chen, S., Chen, J., Zhou, S., Wang, B., Han, S., Su, C., Yuan, Y., Wang, C.: SIGformer: Sign-aware graph transformer for recommendation. In: Proceedings of the 47th International ACM SIGIR Conference on Research and Development in Information Retrieval, pp. 1274--1284. ACM, New York (2024)

\bibitem{zeng2024popularity}
Zeng, X., Chang, C., Tang, F., Wu, Z., Tang, Y.:
Popularity-aware graph neural network with global context for session-based recommendation.
In: Jin, C., Yang, S., Shang, X., Wang, H., Zhang, Y. (eds.)
Web Information Systems and Applications -- WISA 2024.
LNCS, vol. 14883, pp. 163--171.
Springer, Singapore (2024)

\bibitem{song2018nodeedge}
Song, W., Wang, S., Yang, B., Lu, Y., Zhao, X., Liu, X.:
Learning node and edge embeddings for signed networks.
Neurocomputing \textbf{319}, 42--54 (2018)

\bibitem{song2021hyperbolic}
Song, W., Chen, H., Liu, X., Jiang, H., Wang, S.:
Hyperbolic node embedding for signed networks.
Neurocomputing \textbf{421}, 329--339 (2021)

\bibitem{zhao2018block}
Zhao, X., Chen, H., Liu, X., Tan, X., Song, W.:
Block modelling and learning for structure analysis of networks with positive and negative links.
In: Knowledge Science, Engineering and Management, pp. 396--402. Springer (2018)

\bibitem{song2020ssbm}
Liu, X., Song, W., Musial, K., Zhao, X., Zuo, W., Yang, B.:
Semi-supervised stochastic blockmodel for structure analysis of signed networks.
Knowledge-Based Systems \textbf{195}, 105714 (2020)

\bibitem{liu2025sbmsurvey}
Liu, X., Song, W., Musial, K., Li, Y., Zhao, X., Yang, B.:
Stochastic block models for complex network analysis: A survey.
ACM Transactions on Knowledge Discovery from Data \textbf{19}(3), 55:1--55:35 (2025)

\end{thebibliography}
\end{document}